\documentclass[fleqn,twoside]{article}
\usepackage{espcrc2}

\usepackage{graphicx}

\newcommand{\AmS}{{\protect\the\textfont2
  A\kern-.1667em\lower.5ex\hbox{M}\kern-.125emS}}

\title{Precision Determination of $|V_{ub}|$}

\author{G. Paz\address{Institute for Advanced Study, Einstein Drive,
Princeton, NJ 08540, USA }}

\begin{document}

\begin{abstract}
The last two years have seen an impressive improvement in the
determination of $|V_{ub}|$, especially from inclusive decays.  The
error on $|V_{ub}|$ measured with inclusive decays was reduced from
18\% (PDG 2004) to 8\% (PDG 2006). This progress is a result of
combined experimental and theoretical efforts. In this talk, the
theoretical framework (BLNP) that enabled such progress is reviewed,
as well as other approaches to an inclusive determination of
$|V_{ub}|$ (DGE, $M_X-q^2$ etc.). The prospects of improving
$|V_{ub}|$ are discussed, addressing issues of weak annihilation,
implications of leptonic B decays, and determination of $|V_{ub}|$
with exclusive decays.

\vspace{1pc}
\end{abstract}
\maketitle
\section{INTRODUCTION}
$V_{ub}$, one of the smallest matrix element of the CKM matrix, is one
of the fundamental parameters of the Standard Model. In the
geometrical picture of the unitarity triangle, the side opposite to
the angle $\beta\,(\equiv \phi_1)$ is proportional to
$|V_{ub}/V_{cb}|$. Since this angle is measured to high accuracy by
the $B$ factories, and the error on $|V_{cb}|$ is at a level of 2\%
\cite{Yao:2006px}, an accurate measurement of $|V_{ub}|$ is important
for constraining the unitarity triangle.

$|V_{ub}|$ can be measured either through exclusive decays,
e.g. $\bar{B}\to \pi\, l^- \bar\nu$, or inclusive decays: $\bar B\to
X_u\, l^- \bar\nu$. In the former method we encounter a large
theoretical uncertainty as a result of our limited knowledge of the
form factors that control the decay. The latter method offers, in
principle, the most accurate way to extract $|V_{ub}|$ from the total
width $\Gamma\left(\bar B\to X_u\, l^- \bar\nu\right)$. In practice,
since $|V_{cb}|\gg |V_{ub}|$, the total width cannot be measured due
to the large charm background from $\bar B\to X_c\, l^- \bar\nu$
decays. In order to eliminate the charm background one is forced to
look at regions of phase space where charm cannot be produced. In
these regions the theoretical description is more complicated, but
thanks to recent progress in our understanding of charmless inclusive
$B$ decays, the inclusive measurement of $|V_{ub}|$ gives a smaller
error compared to the exclusive one.

The recent global analysis of the unitarity triangle shows a
``tension'' at the level of $2\sigma$ between $\beta\,(\equiv \phi_1)$
and $|V_{ub}|$ \cite{CKMfitter} (see the contribution by S. T'Jampens
at this conference). This is a result of two effects. One, the central
value of $|V_{ub}|$ has increased, especially as measured from
inclusive decays, and the error bar decreased from 18\% in 2004
\cite{Eidelman:2004wy} to 8\% in 2006 \cite{Yao:2006px}.  At the same
time the value of $\sin 2\beta (\equiv \sin 2\phi_1)$ has
decreased. It is therefore important to understand how the value of
$|V_{ub}|$ is obtained.

In this talk we will therefore focus on the theory of inclusive
measurement of $|V_{ub}|$. For various experimental issues we refer
the reader to the contribution by L. Gibbons at this conference, and
for exclusive measurement using lattice data we refer to the
contribution of C. Davies.
\section{KINEMATICS}
We begin our discussion by shortly reviewing the kinematics of
semileptonic $B$ decays.  For a ``pedestrian'' introduction to
semileptonic $B$ decays see chapter 1 of \cite{Paz:2006me}.

Any $\bar B\to X_u\, l^- \bar\nu$ event can be described by three
kinematical variables. The triple differential decay rate depends, via
the optical theorem, on the hadronic tensor $W_{\mu\nu}$ (for
derivation see \cite{Paz:2006me}). Given a basis of two four vectors,
the hadronic tensor is usually decomposed into the various possible
Lorentz structures, where the coefficients of these structures are
called structure functions and denoted as $W_i$.

In \cite{Lange:2005yw} the use of the $(v,n)$ basis was advocated. In
this basis $v$ is the four velocity of the decaying $B$ meson and $n$
is a light-like vector in the direction of the hadronic jet. In the
rest frame of the decaying meson we have $v=(1,0,0,0)$ and we can
choose $n$ to be $n=(1,0,0,1)$. One can also define a conjugate light
like vector $\bar n=(1,0,0,-1)$, such that $2v=n+\bar n$ (in the
following we assume these values for $v,n$ and $\bar n$). This choice
of vectors motivates the following choice of kinematical variables:
\begin{eqnarray}
&& P_l=M_B-2E_l, \nonumber\\
&&P_-=\bar n\cdot P_X=E_X+|\vec{P}_X|,\nonumber\\
&&P_+=n \cdot P_X=E_X-|\vec{P}_X|,
\end{eqnarray}
where $P_X$ is the four momentum of the hadronic jet and $E_l$ is the
energy of the charged lepton. Other kinematical variables can be
expressed in terms of this choice of variables. For example the
hadronic and leptonic invariant masses are: $M_X^2=P_+P_-$, and
$q^2=(M_B-P_-)(M_B-P_+)$, respectively.

In terms of these variables the {\em exact} expression for the triple
differential decay rate is:
\begin{eqnarray}
\label{triple}
&&\frac{d^3 \Gamma_u}{dP_+ \,dP_-\, dP_l} =\frac{G_F^2 |V_{ub}|^2}{16
\pi^3}\; (M_B-P_+) \nonumber\\
&&\Big[ (P_- - P_l)(M_B-P_- +P_l - P_+)\; \tilde
W_1\nonumber\\
&&+(M_B-P_-)(P_--P_+)\; \frac{\tilde W_2}{2} +\nonumber\\
&&(P_--P_l)(P_l-P_+)\; \left(\frac{y}4\tilde W_3 +\tilde W_4+\frac 1 {y}
\tilde W_5\right)\Big],\nonumber\\
\end{eqnarray}
where $y=(P_- -P_+)/(M_B-P_+)$ and $\tilde W_i$ are defined in
\cite{Lange:2005yw} and do not depend on $P_l$. This choice of
variables and basis has two main advantages. The first is that the
phase space has probably the simplest form possible:
\begin{equation}
\frac{M_\pi^2}{P_-} \le P_+ \le P_l \le P_- \le M_B.
\end{equation}
The second advantage is that there is no explicit dependence on the
mass of the $b$ quark in the expression for the triple rate (\ref
{triple}). This allows for theoretical predictions of partial rates
instead of event fractions and eliminates the large source of
uncertainty from the value $m_b$. The triple rate depends on $m_b$
only through the $\tilde W_i$ functions, which we now discuss.
\section{DYNAMICS}
The structure functions $\tilde W_i$ cannot be calculated
exactly. Fortunately for heavy meson decays there are two small
parameters we can expand $W_i$ in: the mass of the $b$ quark (or more
exactly $\Lambda_{\rm QCD}/m_b$) and $\alpha_s$.

If we had no charm background we could integrate over $P_\pm$ up to
$M_B$ and use a Heavy Quark Effective Theory (HQET) based Operator
Product Expansion (OPE) to write $\tilde W_i$ as:
\begin{equation}
\label{OPE}
\tilde W_i \sim c_0 \langle O_0\rangle+c_2 \frac{\langle
O_2\rangle}{m_b^2}+c_3 \frac{\langle O_3\rangle}{m_b^3}+\cdots.
\end{equation}
(We use a short hand notation in here. In practice there are several
operators at each order and the coefficients are generalized functions
and not constants).  The coefficients $c_i(\mu)$ contain the short
distance physics ($\mu\sim m_b$) and are calculable in perturbation
theory ($\langle O_1\rangle=0$ as a result of HQET equations of motion
\cite{Chay:1990da}). Currently $c_0$ \cite{DeFazio:1999sv} is known at
${\cal O}(\alpha_s)$ while $c_2$ \cite{Blok:1993va,Manohar:1993qn} and
$c_3$ \cite{Gremm:1996df} are known at ${\cal
O}(\alpha_s^0)$. Recently even $c_4$ was calculated at ${\cal
O}(\alpha_s^0)$ \cite{Dassinger:2006md}.

The matrix elements of the local operators $O_i$ between $\bar B$
meson states, $\langle O_i\rangle$, are called the Heavy Quark (HQ)
parameters. They contain the long distance physics and must be taken
from experiment. We expect them to scale as $\langle O_i\rangle
\sim\Lambda_{\rm QCD}^i $.  We have $\langle O_0\rangle=1$; $\langle
O_2\rangle$ defines two HQ parameters $\mu^2_\pi$ and
$\mu^2_G=3[(M_B^*)^2-(M_B)^2]/4$; $\langle O_3\rangle$ defines
$\rho_{LS}^3$ and $\rho_D^3$ etc.

A similar OPE can be constructed for $\bar B\to X_c\, l^- \bar\nu$ and
$\bar B \to X_s \gamma$, which contains the {\em same} HQ
parameters. For $\bar B\to X_c\, l^- \bar\nu$ the OPE works very
well. As a result the error on $|V_{cb}|$ is at the level of 2\%
\cite{Yao:2006px} and the HQ parameters can be extracted from
experiment. For $\bar B \to X_s \gamma$ a local OPE is more
problematic for two reasons. The first is that the OPE is valid for
photon energies that cannot be attained at the current
experiments. The second is that even for such energies when one goes
beyond leading order in $1/m_b$ non-local operators arise (see
\cite{Lee:2006wn} and references within).

In practice we have to deal with a large charm background from $\bar
B\to X_c\, l^- \bar\nu$.  In order to eliminate this background
experimental cuts must be imposed. The ``charmless'' region is
typically the region of phase space where $P_+\sim\Lambda_{\rm QCD}$
and $P_-\sim m_b$ . In this region, since $P_+\ll P_-$, the HQET based
OPE is not valid (a fact commonly referred to as ``breakdown of the
OPE''). This region is known as the shape-function (SF) region, to
distinguish it from the region of $P_+\sim P_-\sim m_b$, the OPE
region.

Still, we do have a systematic $1/m_b$ expansion in this region. The
$\tilde{W}_i$ can be written as:
\begin{equation}
\tilde W_i\sim H_u\cdot J\otimes S+\frac 1 {m_b}\sum_k h_u^k \cdot
j_u^k\otimes s_u^k+\cdots.
\end{equation}
The $H,J,$ and $S$'s are called the hard, jet, and shape functions
respectively, and they encode the physics at the scales $\mu_h\sim
m_b$, $\mu_i\sim\sqrt{m_b\Lambda_{\rm QCD}}$, and
$\mu_0\sim\Lambda_{\rm QCD}$, respectively. The structure functions
are sensitive to the ``intermediate'' scale
$\mu_i\sim\sqrt{m_b\Lambda_{\rm QCD}}$ since this was introduced as a
result of the experimental cuts. The hard and jet functions are
calculable in perturbation theory, while the shape functions are
non-perturbative objects. Currently the leading order hard function is
known at ${\cal O}(\alpha_s)$ \cite{Bauer:2003pi,Bosch:2004th} and the
leading order jet function at order ${\cal O}(\alpha_s^2)$
\cite{Becher:2006qw}.  The subleading functions are only known at
${\cal O}(\alpha_s^0)$ \cite{Lee:2004ja,Bosch:2004cb,Beneke:2004in}.

For the photon spectrum in $\bar B \to X_s \gamma$ near the
kinematical end-point, which is the region measured well by
experiments, we have a similar expansion:
\begin{equation}
\frac{d\Gamma}{dE_\gamma}\sim H_s\cdot J\otimes S+\frac {1} {m_b}\sum_k
h_s^k \cdot j_s^k\otimes s_s^k+\cdots.
\end{equation}
Notice that the leading order jet and shape function are the same as
$\bar B\to X_u\, l^- \bar \nu$, while the hard function is
different. Beyond leading order this is no longer the case.

What is the relation between the two regions? Neglecting $\alpha_s$
corrections, moments of the shape functions are related to the HQ
parameters. For example, the first moment of $S$, the leading order
shape function, is related to $m_b$, while the second moment is
related to $\mu_\pi^2$ \cite{Neubert:1993ch}. Including $\alpha_s$
corrections, we can define a renormalization scheme, called the
``shape function scheme" \cite{Bosch:2004th}, such that these
relations hold at each order in perturbation theory. Currently, these
relations are known at order $\alpha_s^2$ for $m_b$ and $\mu_\pi^2$
\cite{Neubert:2004sp}. The high precision of these relations implies
that a good knowledge of the HQ parameters helps to constrain the
shape functions. There are similar relations between the subleading
shape functions and the HQ parameters, although they are known only at
order $\alpha_s^0$. Still, such relations help us to model the
subleading shape functions.

This concludes our brief review of the dynamics of inclusive $B$
decays. We should emphasize that the above description is not a
theoretical model, but a rigorous theory based on QCD and a systematic
expansion in $1/m_b$. Any inclusive extraction of $|V_{ub}|$ has to be
based on these ingredients.

\section{INCLUSIVE EXTRACTION OF \boldmath $V_{ub}$ - PRESENT}
There are currently several theoretical calculations, already
implemented by experiments, which combine some or all of the
ingredients described in the previous section. We now review each of
them briefly (the LLR approach will be discussed in the next section).
For a more detailed account see the original papers.

{\bf BLNP Approach}: The BLNP (Bosch-Lange-Neubert-Paz) approach
\cite{Lange:2005yw} is a culmination of a research efforts that
extends over 12 years. Its ``philosophy'' is to use all that is
currently known about the triple differential decay rate of $\bar B\to
X_u\, l^- \bar\nu$ and $\bar B \to X_s \gamma$, namely
\begin{itemize}
\item  At leading order in $1/m_b$: $H_u,\,H_s,\, J$ at ${\cal O}(\alpha_s)$;
\item  $1/m_b$ subleading shape functions at ${\cal O}(\alpha_s^0)$;
\item Known $\alpha_s/m_b$ terms from the OPE calculation (the part of
$c_0$ in equation (\ref{OPE}) that becomes subleading in the SF
region);
\item Known $1/m_b^2$ terms from the OPE calculation (part of $c_2$ in
equation (\ref{OPE})).
\end{itemize}
The various ingredients are implemented in such a way that there is a
smooth transition between the SF and OPE regions, in the sense that
once the kinematical variables are integrated far enough the OPE
result is recovered.

Non-perturbative physics effects are contained in the leading and
subleading order shape functions. BLNP uses the fact that the leading
order shape function is universal, so the formalism allows the
extraction of the leading order shape function from the photon
spectrum in $\bar B \to X_s \gamma$ and its use as an input for $\bar
B\to X_u\, l^- \bar\nu$ (beyond leading order a slight redefinition of
of $S$ is needed, see \cite{Lange:2005yw}). The subleading shape
functions are modeled in accordance with their moment constraints.

The complete error analysis consists of the following:
\begin{itemize}
\item The leading order shape function is taken from experiment as
explained above, substantially reducing the associated error.
\item The perturbative error is estimated by varying the scale of
$\alpha_s$ in the different terms.
\item The subleading shape function error is estimated as follows. At
order $\alpha_s^0$, there are 3 subleading functions and 9 models are
constructed for each of them. The $9^3=729$ possible combinations are
then scanned to estimate the resulting error. The models for the
subleading shape functions are constructed such that they respect the
moment constraints (the zeroth moment vanishes, the first moment is
related to $\mu^2_{\pi}$ and $\mu^2_G$, and the second moment is of
order $\Lambda^3_{\rm QCD}$), and that the dimensionless shape
functions are ${\cal O}(1)$ in a region where their arguments are of
order $\Lambda_{\rm QCD}$.
\item The weak annihilation error (see below) is taken as a fixed
percentage of the total rate.
\end{itemize}
The HFAG average for ICHEP 2006 \cite{HFAG} gives the value
$|V_{ub}|=(4.49\pm0.19\pm0.27)\cdot 10^{-3}$ using the BLNP approach,
where the first error is from experiment and the second from
theory. The non experimental errors associated with this result are:
4.2\% HQ error (for the leading order shape function), 3.8\% for the
combined perturbative and subleading shape function error and 1.9\%
for weak annihilation.

{\bf BLL Approach}: The BLL (Bauer-Ligeti-Luke) approach
\cite{Bauer:2001rc} is based on the fact that if a low $q^2$ cut is
imposed to eliminate charm background, an OPE expansion for the
partial rate can be constructed, which is suppressed by inverse powers
of $m_c$ instead of $m_b$. (This is possible since for such a cut
$P_+\sim P_-\sim m_c$). In order to optimize both the efficiency and
the theoretical uncertainty, the authors of \cite{Bauer:2001rc}
suggested to use a combined $M_X-q^2$ cut. The main ingredients of the
BLL approach are:
\begin{itemize}
\item The OPE is assumed to be valid for the combined cut. 
\item The LO shape function sensitivity is estimated by convoluting
the tree-level decay rate with the difference between the
``tree-level'' shape function model and a delta function model for the
shape function.
\item The subleading shape function contribution is assumed to be small
and is not assessed.
\end{itemize}
The HFAG average for ICHEP 2006 \cite{HFAG} gives the value
$|V_{ub}|=(5.02\pm0.26\pm0.37)\cdot 10^{-3}$ using the BLL approach,
where the first error is from experiment and the second from
theory. The theory error includes a $3\%$ contribution from shape
function sensitivity.

{\bf DGE Approach}: The Dressed Gluon Exponentiation (DGE) approach is
advocated by Andersen and Gardi \cite{Andersen:2005mj}. Similar to the
BLNP approach it can be applied to various experimental cuts, but
conceptually it is different from the other approaches discussed in
this talk.  For a less technical review of the DGE approach we refer
the reader to \cite{Gardi:2006jc}. Here we only mention the following
features of this approach:

\begin{itemize}
\item The decay spectra is approximated by the resummed on-shell
$b$-quark decay spectrum, for which the only input parameters are
$m_b$ and $\alpha_s$.
\item Non-perturbative effects associated with the meson structure are
estimated using renormalon analysis.
\end{itemize} 

The HFAG average for ICHEP 2006 \cite{HFAG} gives the value
$|V_{ub}|=(4.46\pm0.20\pm0.20)\cdot 10^{-3}$ using the DGE approach,
where the first error is from experiment and the second from theory.
 
{\bf Discussion:} Let us compare the various approaches. As might be
expected, the main difference between them is the estimate of non
perturbative effects and power corrections, beginning with the leading
order shape function.

In the BLNP approach the leading order shape function is to be
extracted from experiment. In the current experimental implementation
the full shape of the photon spectrum is not used and only the first
two moments of the shape function are used to constrain its form. One
might argue that the possible variations of the photon spectrum are
already included in the range of the HQ parameters, but a verification
of this assumption would be desirable. In the BLL approach the shape
function sensitivity was included as described above. We should note
that the BLNP analysis did not find reduced shape function sensitivity
for a combined $M_X-q^2$ cut. Considering the theoretical advances in
the control over the leading order shape function a reevaluation of
this issue in the BLL approach is in order. In the DGE approach there
is no error associated with the leading order shape function. Instead
there is an error from the value of $m_b$, $\alpha_s$ and the
parameter $C$ defined in \cite{Andersen:2005mj}. (Recently the effects
of the non perturbative parameter $f^{\rm PV}$ on $\bar B \to X_s
\gamma$ were also considered \cite{Andersen:2006hr}).  It is unclear
how this $C$ parameter is related to the shape function uncertainty in
all the other approaches, or to the shape function moments. It is also
unclear how in the DGE approach the OPE result is recovered beyond
leading order in $1/m_b$, e.g. the terms proportional to $\mu^2_\pi$
and $\mu^2_G$, when one integrates over the kinematical variables.

Apart from the leading order shape function, we have the issue of
subleading shape functions. In the BLNP approach the subleading shape
were modeled using their moments. In the BLL approach the subleading
shape functions contribution was assumed to be small and was not
assessed. In the DGE approach there is no error from subleading shape
functions. Again, it is unclear how the parameters $C$ and $f^{\rm
PV}$ in the DGE approach are related to the subleading shape function
error in the other approaches.

Despite all of these differences, the inclusive measurement are all
consistent with each other, even if we assume that the error bars are
underestimated. We have no good explanation for this fact.
\section{IMPROVED \boldmath $V_{ub}$}
The extraction of $|V_{ub}|$, impressive as it is, can be further
improved. We separate the discussion of improvements that can be
implemented today, using the currently available theoretical tools,
and future feasible improvements.

{\bf Improved \boldmath $V_{ub}$ - Today}: In the former class we have
the treatment of weak annihilation (WA) and the weight function
approach.

Weak annihilation appears at order $1/m_b^3$ in the OPE. It arises
from flavor specific four quark operators of the form $\bar b\,\Gamma
u\,\bar u\,\Gamma b$, and effects neutral and charged $B$'s
differently \cite{Bigi:1993bh}. Currently there are only estimates of
its magnitude \cite{NPS}. For example a CLEO analysis finds the limit
$\Gamma_{\rm WA}/\Gamma_{b\to u}<7.4\%$ at the 90\% confidence level
\cite{Rosner:2006zz}. Apart from estimating the WA as a fixed
percentage of the rate, a different strategy was suggested by Lange,
Neubert and Paz \cite{Lange:2005yw}. Since this conference took place
in Oxford UK, we can follow the Queen of Hearts and summarize this
strategy as: ``Off with its head!'' \cite{Carroll:1865}. More
concretely, since weak annihilation is concentrated in the region of
$q^2=m_b^2$, if we cut on {\em high} $q^2<q^2_{\rm max}$
(e.g. $q^2_{\rm max}=(M_B-M_D)^2$) combined with a $M_X$ or $P_+$ cut
to eliminate charm background, we would eliminate the WA error. With
such a cut one loses efficiency but this might be compensated by the
elimination of the WA error. Preliminary studies in
\cite{Lange:2005yw} showed that this is indeed the case. {\em This
method is still waiting for experimental implementation}.

The weight function idea is to directly relate the photon spectrum of
$\bar B \to X_s \gamma$ to $\bar B\to X_u\, l^- \bar\nu$ spectra
without a need to extract the leading order shape function, and was
first suggested in \cite{Neubert:1993um}. The theoretical input in
this approach is the weight function, which at leading order can be
calculated in perturbation theory:
\begin{equation}
W\sim \frac{\Gamma_u}{\Gamma_s}\sim \frac{H_u\cdot
J[m_by(P_+-\hat{\omega})]\otimes S(\hat{\omega})+...}{H_s\cdot
J[m_b(P_+-\hat{\omega})]\otimes S(\hat{\omega})+...}.
\end{equation}
In this relation the leading order shape function cancels in the
ratio. Beyond leading order this is no longer the case. BaBar has used
a calculation of a weight function for the $M_X$ spectrum by
Leibovich, Low, and Rothstein (LLR) \cite{Leibovich:2000ey} to measure
$|V_{ub}|$. The result is $|V_{ub}|=(4.43\pm0.45\pm0.29)\cdot 10^{-3}$
\cite{Aubert:2006qi}, where the first error is from experiment and the
second from theory.

A more recent theoretical calculation was performed by Lange, Neubert,
and Paz \cite{Lange:2005qn}. In this approach a weight function that
relates the {\em normalized} photon spectrum in $\bar B \to X_s
\gamma$ to the $P_+$ in $\bar B\to X_u\, l^- \bar\nu$ is
constructed. The main reason to use the normalized spectrum is the
better perturbative convergence of the weight function. This weight
function contains two loop corrections from the ratios of jet
functions, subleading shape function corrections and the known
$\alpha_s/m_b$ corrections. The combined theoretical error is at a
level of 5\%. The approach was generalized by Lange
\cite{Lange:2005xz} and weight functions were calculated for an
arbitrary $\bar B\to X_u\, l^- \bar\nu$ spectra. These weight
functions have the potential to give the best extraction of
$|V_{ub}|$. {\em This method is still waiting for experimental
implementation}.

{\bf Improved \boldmath $V_{ub}$ - Future:} Apart from these already
available calculations we can construct a wish list for feasible
theoretical calculations. First, a complete subleading shape function
analysis for $\bar B \to X_s \gamma$ is underway \cite{inprep} and
preliminary results were already reported in \cite{Lee:2006wn}. With
the approaching completion of the full order $\alpha_s^2$ corrections
to $c_0$ for $\bar B \to X_s \gamma$ \cite{Misiak:2006zs} (see also
the contribution by T. Hurth at this conference) the calculation of
$H_s$ at ${\cal O}(\alpha_s^2)$ is nearly done
\cite{Becher:2006pu}. In order to construct a full two loop weight
function a two loop expression for $H_{u}$ is needed which is not easy
but feasible. The next source of uncertainty is the $\alpha_s$
corrections for the terms containing the subleading shape
functions. Such a calculation is more complicated and connected to the
calculation of $c_{2}$ in equation (\ref{OPE}) to order $\alpha_s$,
where the latter would serve as a check to the former. Finally it is
not completely clear that the subleading shape functions cannot be
extracted from data. With the completion of this wish list we would
probably reach the boundary of the theoretical accuracy in extracting
$|V_{ub}|$.

\section{EXCLUSIVE \boldmath $|V_{ub}|$ AND LESSONS FROM LEPTONIC $B$ DECAYS}

$|V_{ub}|$ can also be extracted from exclusive decays such as
$B\to\pi l\bar \nu$ (see the contribution by L. Gibbons at this
conference). In order to do so there is a need for a theoretical input
about the form factor $f_+(q^2)$. Currently there are two different
theoretical approaches. The first being lattice QCD calculations,
valid for $q^2>16$ GeV, and the second light cone sum rules valid for
$q^2<16$ GeV. For the first approach HFAG \cite{HFAG} cites three
sources: an unquenched calculation by HPQCD collaboration with
$|V_{ub}|=(3.93\pm0.26^{+0.59}_{-0.41})\cdot 10^{-3}$
\cite{Dalgic:2006dt} (see also the contribution by C. Davies at this
conference), an unquenched calculation by the FNAL collaboration with
$|V_{ub}|=(3.51\pm0.23^{+0.61}_{-0.4})\cdot 10^{-3}$
\cite{Okamoto:2004xg}, and a quenched calculation which can be found
in \cite{HFAG}. For the second approach there is only a single source:
a calculation by Ball and Zwicky \cite{Ball:2004ye} that gives
$|V_{ub}|=(3.38\pm0.12^{+0.56}_{-0.37})\cdot 10^{-3}$.  As can be seen
from the above list, the exclusive predictions are typically smaller
than the inclusive ones, which was also ``historically'' the case
\cite{Eidelman:2004wy,Hagiwara:2002fs}.

Recently the Belle collaboration reported on evidence for the (pure)
leptonic $B\to \tau\bar{\nu}_\tau$ decay \cite{Ikado:2006un}. The
branching fraction of such a decay depends on the product of
$|V_{ub}|$ and $f_B$, the $B$ meson decay constant. The product was
measured to be $f_{B} \cdot |V_{ub}| =
(10.1^{+1.6}_{-1.4}(\mbox{stat})^{+1.3}_{-1.4}(\mbox{syst})) \times
10^{-4}$ GeV. Using the inclusive value $|V_{ub}|=(4.39\pm 0.33)\cdot
10^{-3}$ the decay constant can be extracted: $f_{B} =
0.229^{+0.036}_{-0.031}(\mbox{stat})^{+0.034}_{-0.037}(\mbox{syst})$.
This value is in good agreement with the unquenched lattice value:
$f_B = 0.216\pm 0.022$ GeV \cite{Gray:2005ad}, or the QCD sum rule
calculations $f_B = 0.210\pm 0.019$ GeV \cite{Jamin:2001fw} and $f_B =
0.206\pm 0.020$ GeV \cite{Penin:2001ux}, supporting the value of the
inclusive $|V_{ub}|$.

\section{CONCLUSIONS}

The last two years have seen an impressive improvement in the
determination of $|V_{ub}|$, especially from inclusive decays.  This
is a result of combined experimental and theoretical hard work. In
this talk we have reviewed some of the theoretical work. Further
improvement is also possible. As we have pointed out, there are
theoretical tools that still await experimental implementation, namely
a cut on high $q^2$ to eliminate weak annihilation, and advanced two
loop relations between the photon spectrum in $\bar B \to X_s \gamma$
and $\bar B\to X_u\, l^- \bar\nu$ spectra. Beyond these, more
theoretical improvement is also feasible.

The time has now come for a critical comparison of the theoretical
approaches to inclusive $|V_{ub}|$, namely, comparing the underlying
assumptions, the perturbative and non perturbative corrections. This
is especially important if we want to take seriously the $2\, \sigma$
``tension'' between $\sin 2\beta (\equiv \sin 2 \phi_1)$ and inclusive
$|V_{ub}|$

\section*{ACKNOWLEDGMENTS}
I would like to thank the organizers of Beauty 2006 for a inviting me
to give this talk and for a highly enjoyable conference. I also would
like to thank Stefan Bosch, Bjorn O. Lange, and Matthias Neubert. Much
of the work presented here was done in collaboration with
them. Finally I would like to thank Einan Gardi, Michael Gronau, Jon
Rosner, Stephane T'Jampens, and Guy Wilkinson for useful
discussions. This work was supported in part by the Department of
Energy under Grant \#DE-FG02-90ER40542 and by the United States-Israel
Binational Science Foundation grant \#2002272.

\end{document}